\documentclass[10pt,letterpaper,twocolumn]{article}
\usepackage{ol2}
\usepackage{amsmath}

\usepackage{graphicx}

\newcommand\leqt[1]{\protect\label{eq:#1}}
\newcommand\reqtn[1]{\ref{eq:#1}}
\newcommand\reqt[1]{(\reqtn{#1})}

\begin{document}
\twocolumn[ 
\title{Spatial-Spectral Vortex Solitons in Quadratic Lattices}

\author{Zhiyong Xu$^{*}$ and Andrey A. Sukhorukov}

\affiliation{Centre for Ultra-high bandwidth Devices for Optical Systems (CUDOS),
Nonlinear Physics Centre, Research School of Physics
and Engineering, The Australian National University, Canberra ACT, 0200, Australia \\
$^*$Corresponding author: xzy124@physics.anu.edu.au}

\begin{abstract}
We predict the existence of spatial-spectral vortex solitons in one-dimensional periodic waveguide arrays with quadratic nonlinear response. In such vortices the energy flow forms a closed loop through the simultaneous effects of phase gradients at the fundamental frequency and second-harmonic fields, and the parametric frequency conversion between the spectral components. The linear stability analysis shows that such modes are stable in a broad parameter region.
\end{abstract}

] 

\maketitle
Manipulation of light beams and pulses in nonlinear photonic lattices or waveguide arrays is attracting increasing attention, due to the potential to control spatial beam shaping combined with manipulation of temporal and spectral characteristics~\cite{Lederer:2008-1:PRP}. In particular, photonic lattices created in a medium with quadratic nonlinearity can facilitate ultra-fast all-optical switching through
parametric wave mixing between the fundamental and second-harmonic waves~\cite{Pertsch:2003-102:OL}. Various approaches to beam manipulation reply on the special features of localized modes in the form of discrete or lattice solitons~\cite{Peschel:1998-1127:PRE, Darmanyan:1998-2344:PRE, Malomed:2002-56606:PRE, Kartashov:2004-1117:OL, Lederer:2008-1:PRP}.

In this Letter, we predict the appearance of a different type of discrete quadratic solitons in one-dimensional lattices. In contrast to previously known solitons, such localized states exhibit directional power flows between the lattice sites, that are compensated through parametric conversion between the fundamental and second-harmonic waves. Since the power flow loop is closed in spatial-spectral domain, we refer to these solutions as spatial-spectral vortices.

The spatial light dynamics in a one-dimensional photonic lattice created in a medium with quadratic nonlinearity can be modelled by a set of coupled equations for the amplitudes of fundamental frequency (FF) and second-harmonic (SH) modes of individual waveguides~\cite{Lederer:2008-1:PRP}, which can be written in the normalized form:
%
\begin{equation} \leqt{DNLS}
   \begin{array}{l} {\displaystyle
       i\frac{d A_n}{dz} + c_1\left( A_{n+1}+A_{n-1}\right) +\gamma A_n^{*}B_n  = 0
    }\\*[9pt]{\displaystyle
       i\frac{d B_n}{dz} + c_2\left( B_{n+1}+B_{n-1}\right) +\beta B_n+\gamma A_n^2
            = 0
    } \end{array}
\end{equation}
where $z$ is the normalized propagation distance along the waveguides, $A_n$ and $B_n$ are the normalized FF and SH mode amplitudes in the $n$th waveguide, respectively. The coefficients $c_{1,2}$ define the linear coupling between the guided modes at the corresponding frequency components. Parameter $\beta$ defines the phase mismatch between the FF and SH modes. The strength of quadratic nonlinearity is characterized by the coefficient $\gamma$, and with no loss of generality it can be scaled to unity ($\gamma=1$).

In order to study vortex states, it is important to consider the mechanisms of power flows. The power density at the lattice site number $n$ can be defined as $|A_n|^2$ and $|B_n|^2$ at the FF and SH spectral components, respectively. Whereas it can be shown that the overall power is conserved, $I=\sum_n\left( \left| A_n\right| ^2+\left| B_n\right| ^2\right) = {\rm const}$, the power density at individual lattice sites can change due to (i)~mode coupling with the neighboring lattice sites of the same frequency component and (ii)~parametric frequency conversion. Indeed following from Eq.~\reqt{DNLS}, we have $d|A_{n}|^{2}/dz= J_{1}(n-1,n) - J_{1}(n,n+1) - J_{p}(n)$, and $d|B_{n}|^{2}/dz = J_{2}(n-1,n) - J_{2}(n,n+1) + J_{p}(n)$, where $J_{1}(n,n+1)=2c_{1}\texttt{Im}(A_{n}^{*}A_{n+1})$ and $J_{2}(n,n+1)=2c_{2}\texttt{Im}(B_{n}^{*}B_{n+1})$ define the `spatial' energy flows between the lattice sites $(n)$ and $(n+1)$ of the FF and SH components, respectively, and $J_{p}(n)=-2\gamma \texttt{Im}(A_{n}^{2}B_{n}^{*})$ defines the `spectral' energy flow between FF and SH components due to quadratic nonlinearity.
%

We seek vortex solutions where the energy flows in spatial and spectral domains form closed loops, such that there is no overall energy redistribution between the waveguides along the propagation direction. Such states correspond to stationary solutions for Eqs.~\reqt{DNLS} in the form $A_n=A_0(n) \exp\left[ i b z + i \varphi_1(n) \right]$, and $B_n=B_0(n) \exp\left[ 2i b z+ i \varphi_2(n) \right]$,
where $b$ is the real propagation constant. After substituting these expressions in Eqs.~\reqt{DNLS}, we obtained a set of stationary nonlinear equations for the real amplitude [$A_0(n)$ and $B_0(n)$] and phase [$\varphi_{1,2}(n)$] profiles of FF and SH components, which were solved with the numerical relaxation technique.
The power flows for stationary solutions are expressed as: $J_1(n,n+1)= 2 c_{1} A_0(n) A_0(n+1) \sin[ \varphi_1(n+1) - \varphi_1(n) ]$, $J_2(n,n+1)= 2 c_{2} B_0(n) B_0(n+1) \sin[ \varphi_2(n+1) - \varphi_2(n) ]$, and $J_{p}(n)= 2 \gamma |A_0(n)|^{2} B_0(n) \sin[ \varphi_2(n) - 2 \varphi_1(n)]$.
Note that the power flows are non-zero only when there appears nontrivial phase difference (not $0$ or $\pi$) at adjacent lattice sites for FF and SH fields.
Whereas complex solutions were identified for quadratic couplers~\cite{Bang:1997-7257:PRE}, only in-phase or twisted localized modes with vanishing power flows were previously found in quadratic lattices~\cite{Peschel:1998-1127:PRE, Darmanyan:1998-2344:PRE, Sukhorukov:2001-16615:PRE, Malomed:2002-56606:PRE}. We also note that complex phase-twisted states were found within the framework of single extended discrete nonlinear Schr\"odinger equation~\cite{Oster:2005-25601:PRE}.

\begin{figure}[htb]
\centerline{\includegraphics[width=6.5cm]{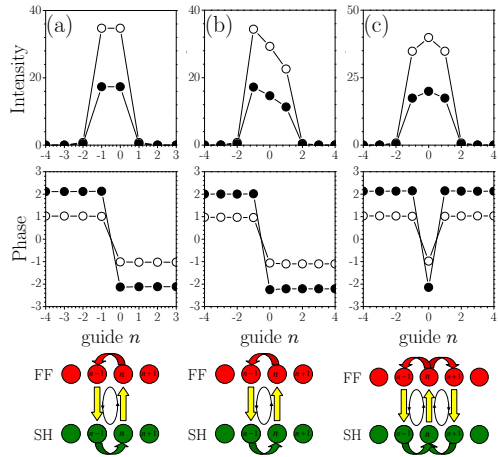}}
\caption{(Color online) Examples of SSV: column (a) symmetric vortex; column (b) asymmetric single-charge vortex, and column (c) vortex with two power flow loops. Here the first and second rows show the intensity and phase distributions of SSV, where lines with white and black circles show the FF and SH fields, respectively. In the third row, a scheme illustrates the energy flow of SSV. Here $\beta=0$, $b=4$, and $c_{2}=2c_{1}=1$.}
\label{figure1}
\end{figure}

We find new classes of solutions featuring nontrivial phase shift between the neighboring lattice sites.
Some illustrative examples of nontrivial phase-twisted modes are presented in Fig.~\ref{figure1}, from which one can see that the stationary solutions are complex, in sharp contrast with the trivial phase twisted modes which have real transverse profiles. Thus we identify them as self-localized optical vortices, the \emph{spatial-spectral vortices} (SSV). For the fundamental vortex solution presented in Fig.~\ref{figure1}(a), the structure of energy flows can be qualitatively explained as follows: (i) the energy flow of the FF wave is directed in one spatial direction (shown by top arrow); (ii) the energy flow is oppositely directed at the SH wave (shown by bottom arrow). One can see clearly that the energy flows between the fundamental and second-harmonic waves are counter-directed at different spatial locations (shown by vertical arrows).
We also find a rich variety of different vortex states. The solution presented in Fig.~\ref{figure1}(b) has an asymmetric shape with one loop of energy flow, which can be viewed as asymmetric SSV. The mode shown in Fig.~\ref{figure1}(c) features two loops of the energy flow.
We find that the SSV presented in Figs.~\ref{figure1}(a) and~(c) have similar properties, whereas asymmetric SSVs are found to be unstable. Therefore, below we present the comprehensive analysis of existence and stability properties of SSV with a single energy flow loop, as shown in Fig.~\ref{figure1}(a).

\begin{figure}[htb]
\centerline{\includegraphics[width=6.5cm]{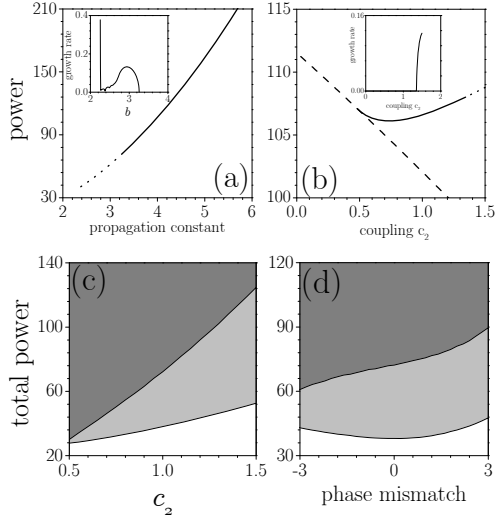}}
\caption{Existence and stability of SSV as shown in Fig.~\ref{figure1}(a):
(a,b)~Power and perturbation growth rate (inset) vs. (a)~the propagation constant at $c_{2}=1$,  $\beta=0$ and
(b)~the coupling coefficient of SH field at $b=4$, $\beta=0$. Solid lines correspond to stable SSV, dotted - unstable SSV, and dashed - twisted solitons without vortex flows.
(c,d)~Stable (gray) and unstable (light gray) domains versus (c)~the coupling coefficient of SH field at $\beta=0$, and (d)~the phase mismatch at $c_{2}=1$. In all the cases $c_{1}=0.5$.}
\label{figure2}
\end{figure}

For a given waveguide array characterized by specific values of coefficients in model Eqs.~\reqt{DNLS}, the vortex solutions form families parametrized by the propagation constant $b$. An important characteristic of SSV is the total power, as shown in Fig.~\ref{figure2}(a). There exists a cutoff value of propagation constant $b$, which depends on the sign and absolute value of phase mismatch parameter, and the coupling coefficients. We also find that the power is a nonmonotonic function of the propagation constant, and there is a narrow region near cutoff (not even visible in the figure) where $dI/db<0$. Our calculations show that the SSV have the power which is approximately two times higher than the power of fundamental (single-site) discrete solitons. Therefore such SSV can be considered as nontrivial bound states of fundamental solitons.

The key point of this work is that SSV appear due to the simultaneous effects of parametric wave mixing and linear mode coupling between the waveguides. We plot the dependence of soliton power on the ratio of coupling coefficients for a fixed propagation constant and phase mismatch parameters in Fig.~\ref{figure2}(b). In this plot the solid and dotted lines corresponds to SSV with complex phase profiles, and we see that the existence of SSV requires that the coupling strength for the SH field exceeds a certain threshold. The SSV are found to bifurcate from the twisted soliton having $0$ or $\pi$ phase jump between lattice sites (shown with dashed line), above a critical value of the SH coupling coefficient.

\begin{figure}[htb]
\centerline{\includegraphics[width=6.5cm]{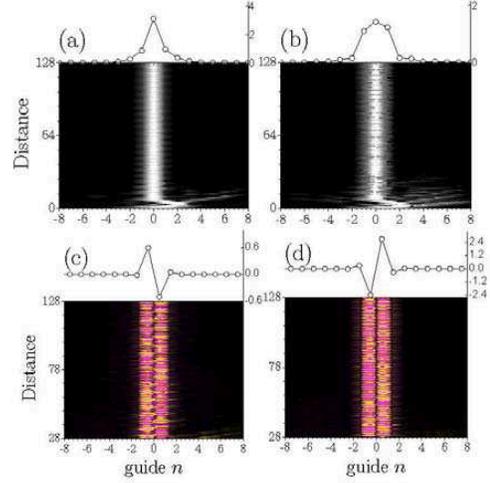}}
\caption{(Color online) Generation of spatial-spectral vortex with FF beam input.
Top row: evolution of amplitude modulus for (a) FF and (b) SH components.
Bottom row: evolution of energy flows for (c) FF and (d) SH components. The insets with circles show the output profiles. Here $\beta=0$, and $c_{2}=2 c_{1}=1$.}
\label{figure5}
\end{figure}

It is important to determine the stability properties of vortices. We perform the linear stability analysis by considering the  perturbed solutions in the form $A_n(z) =\left\{ A_0(n) e^{i \varphi_1(n)}+f_1(n) e^{\delta z}   \right\} e^{ i b z}$, and $B_n(z) =\left\{ B_0(n) e^{i \varphi_2(n)}+f_2(n) e^{ 2 \delta z } \right\} e^{ 2i b z}$.
%
%
Here $\delta$ is the instability growth rate, and $f_{1,2}(n)$ are the corresponding mode profiles. These were determined by numerical solutions of the eigenvalue problem, obtained by linearizing Eqs.~\reqt{DNLS} with respect to small-amplitude perturbations.
We indicate stable SSV with solid lines, and unstable SSV with dotted lines in Figs.~\ref{figure2}(a) and~\ref{figure2}(b), and insets in these figures show the real part of the perturbation growth rate.
Our results show that SSV are stable when their power exceeds a certain threshold [see Fig.~\ref{figure2}(a)], as the growth rate goes to zero. When the power is below a critical value, SSV becomes unstable and suffer from oscillatory instability (leading to its decay into the fundamental solitons). The variation of phase mismatch conditions does not change dramatically the stability-instability scenarios for SSV, and the whole instability domain shown in Fig.~\ref{figure2}(d) confirms that SSV are stable when their power is above a certain value even for different phase mismatch conditions. The coupling strength also affects the stability of SSV. Fig.~\ref{figure2}(b) shows that for a fixed nonlinear wave number shift the SSV become more stable for smaller coupling strength. Eventually when the coupling strength is below the critical value SSV transform into trivial phase-twisted discrete solitons (shown with dashed line), which are stable in the same parameter regime. Note that the instability domain of SSV increases with the increase of the SH coupling for the fixed coupling strength of FF field [Fig.~\ref{figure2}(c)]. The results of linear stability analysis have been confirmed by the extensive numerical simulations of Eq.~\reqt{DNLS} using the beam propagation method.

To verify that the SSV can be observed in experiment, we simulate their generation under practical conditions. We consider a $7\texttt{cm}$ long waveguide array formed by titanium in-diffusion into the Z-cut LINBO$_{3}$ surface, and the interchannel half-beat coupling length for FF wave is $L_{c}=22\texttt{mm}$, which results in the coupling strength for FF wave of $c_{1}=0.5$. The phase-mismatch can be changed by tuning the sample temperature $T$, which is given by $\beta=-\triangle k L=8.1(234-T[^{0}C])$. Additionally, strong linear coupling between the SH waveguide modes is required (we assume $c_{2}=2 c_{1}$). Although in previous experiments~\cite{Iwanow:2004-113902:PRL} $c_2 \ll c_1$, strong coupling can be achieved for higher-order SH guided modes, which can be selectively excited when their propagation constants satisfy the phase-matching condition~\cite{Amoroso:2003-443:IPTL}.
We consider FF wave as an input, and launch a Gaussian beams with a phase tilt along the sites, thus the input FF beam is in the form $A_{n}(z=0)=a \texttt{exp}[-((x_{n}+x_{0})/r_{0})^{2}]\texttt{exp}(i \alpha n)$, where $a$ being amplitude, $x_{0}=0.5$ is the transverse dislocation, and $\alpha$ being the phase tilt. Fig.~\ref{figure5} shows an illustrative example for the excitation of SSV. Indeed, even as the stationary state is reached, the power flows between the lattice sites persist.
The estimated input power required for the vortex generation is $2\texttt{KW}$ (corresponds to $a=3.5$, $r_{0}=3.0$, and $\alpha=\pi/4$).

In conclusion, we have shown that spatial-spectral vortices can appear in quadratic nonlinear waveguide arrays. Such modes have a nontrivial phase between neighboring sites, facilitating non-vanishing energy flows along closed loops.
We stress that the existence of SSV is only allowed due to the effect of parametric wave mixing. These results suggest new opportunities in using nonlinear processes to control the flow of light.

The authors acknowledge fruitful discussions with Prof. Yuri Kivshar. This work has been supported by the Australian Research Council.

\end{document}